\documentstyle[11pt,aaspp4]{article}

\slugcomment{To appear in The Astrophysical Journal}

\lefthead{Gonzalez}
\righthead{Malmquist Bias and the Distance to the Virgo Cluster}

\begin{document}

\title{Malmquist Bias and the Distance to the Virgo Cluster}

\author{Anthony H. Gonzalez}
\affil{Dept. of Astronomy and Astrophysics, University of California at
Santa Cruz, Santa Cruz, CA 95064, Email: anthonyg@ucolick.org} 
\authoremail{anthonyg@ucolick.org}

\author{S. M. Faber}
\affil{UCO/Lick Observatory}
\affil{Dept. of Astronomy and Astrophysics, University of California at
Santa Cruz, Santa Cruz, CA 95064, Email: faber@ucolick.org} 
\authoremail{faber@ucolick.org}

\begin{abstract}
This paper investigates the impact of
Malmquist bias on the distance to the Virgo cluster determined by
the H$_0$ Key Project using M100, and
consequently on the derived value of $H_0$. 
Malmquist bias is a volume-induced statistical effect which causes the
most probable distance to be different from the raw distance measured.
Consideration of the bias in the distance to the Virgo cluster raises this
distance and lowers the calculated value of $H_0$. Monte Carlo simulations 
of the cluster have been run for several possible distributions of 
spirals within the cluster and of clusters in the local universe.
Simulations consistent with known information regarding the cluster and the
errors of measurement result in a bias of about 6.5-8.5\%.
This corresponds to an unbiased distance of 17.2-17.4 Mpc and a value of $H_0$
in the range 80-82 km s$^{-1}$ Mpc$^{-1}$.

	The problem of determining the bias to Virgo illustrates several
key points regarding Malmquist bias.  Essentially all conventional
astronomical distance measurements are subject to this bias. In addition, 
the bias accumulates when an attempt is made to construct ``distance ladders"
from measurements which are individually biased. As will be shown in the case
of Virgo, the magnitude and direction of the bias are sensitive to the spatial
distribution of the parent population from which the
observed object is drawn - a distribution which is often poorly known. This
leads to uncertainty in the magnitude of the bias, and adds to the 
importance of minimizing the number of steps in ``distance ladders".
\end{abstract}


\section{Introduction}

One of the longest standing challenges in modern astronomy is
determination of the value of the Hubble constant. Since Edwin Hubble
first observed the relation between distance and velocity, numerous
attempts have been made to accurately determine the Hubble constant
relating these quantities.  A variety of methods has been applied,
but, despite the extensive work, different observations continue to
give values that differ in their extreme by nearly a factor of two
(\cite{ken95}).  The H$_0$ Key Project was designed to address this
situation, its primary goal being ``to provide a measure of the
Hubble constant accurate to 10\%" (\cite{fre94}). With this goal in
mind, the H$_0$ Key Project has chosen to use Cepheid observations as the
cornerstone method of determining
distances to target galaxies. Cepheids provide a direct method
of determining distance and, with HST, allow distance measurements to
galaxies as distant as the Virgo and Fornax clusters. These
observations can be used both to calculate the Hubble constant
directly and to calibrate secondary distance determination methods
that can then be applied to more distant galaxies. 

	Using Cepheids, the Key Project directly determined the
distance to M100 to be $16.1\pm1.3$ Mpc (\cite{fer96}).  To obtain a value
of $H_0$, M100 was assumed to be at the same distance as the center of
the Virgo cluster, with an estimated Gaussian uncertainty of 20\% due to the
extended nature of the cluster.  With this assumption  and using a
velocity for the cluster of $1,396\pm96$ km s$^{-1}$, the Key Project data yields
$H_0=87\pm20$ km s$^{-1}$ Mpc$^{-1}$.

	The goals of this paper are to bring to light the
effect which Malmquist bias can have on distance measurements such 
as this one and secondarily to obtain a distance to the Virgo cluster
corrected for this bias. Considering bias in the distance to Virgo serves
to illustrate a more general point. Raw astronomical distance estimates are 
rarely unbiased; virtually every measured distance needs correction
for Malmquist bias due to observational uncertainty. In the case of
Virgo, it will be seen that the bias should not be ignored
 although
uncertainty in the structure of the Virgo cluster precludes a precise
determination of it.

\section{Malmquist Bias}
	In the literature, the term Malmquist bias has been used in
different places to describe related but different effects (\cite{wil91}).
In the classical sense, the term Malmquist bias refers to a positive bias in
the luminosity of objects in magnitude limited {\it samples}, as originally
described by Malmquist in 1924 (\cite{mal24}). However, the term is
now also commomly used for a related geometric bias that arises in
distance measurements(\cite{lyn88}). Of particular concern in this paper is the
``inferred distance problem"(\cite{wil91}). In this context, to say
that a measured, raw distance is subject to Malmquist bias means that
the most probable value of the true distance is not the same as the raw
distance.

	Malmquist bias is a statistical effect and formally can be
evaluated using probability theory.  The general form of the expression
for the bias to a single object can be written as 
\begin{equation}
(1+B)D=\left<r|D\right>=\frac{\int_{0}^{\infty} r P(r|D) dr
}{\int_{0}^{\infty} P(r|D) dr }=\frac{\int_{0}^{\infty} r P(r,D) dr
}{\int_{0}^{\infty} P(r,D) dr }, 
\end{equation}
where  $B$ is the bias, $D$ is the raw distance, $r$
 is the actual distance to the object being observed,
$\left<r|D\right>$ is the expectation value of $r$ given $D$,    
$P(r|D)$ is the conditional probability of $r$ given $D$, and 
$P(r,D)$ is the joint probability. This joint probability $P(r,D)$ can be
rewritten as 
\begin{equation}
P(r,D)=P(r)P(D|r)=r^2 n(r) P(D|r).   \label{eq:jntprob}
\end{equation}
The $n(r)$ term is a distribution function describing the parent population
from which the object is drawn (in our case it is the distribution of  
Virgo-like clusters as a function of distance), so $r^2 n(r)$ is effectively
the probability of selecting an object at a given distance.
Meanwhile, $P(D|r)$ contains all
information about uncertainties associated with the measurement. If the
only uncertainty is a Gaussianly distributed magnitude error, $\sigma$,
then this is of the form
\begin{equation}
P(D|r) \propto \exp\left(-\frac{(5\ln (r/D))^2}{2\sigma^2(\ln 10)^2}\right).
\end{equation}
For Virgo, there is also uncertainty in the location of M100 relative
to the center of Virgo.

The result for the simple case of a constant $n(r)$ and Gaussian magnitude
error was given by Lynden-Bell et al. (1988):
\begin{equation}
\left<r|D\right>=D \exp(3.5\Delta^2) \approx D (1+3.5\Delta^2), 
	\label{eq:lyndbell}
\end{equation}
for small $\Delta$. Here, $\Delta=\frac{\ln 10}{5} \sigma$, and so the bias is 
proportional to $\sigma^2$. 

Unfortunately, for more complicated $n(r)$ and $P(D|r)$ this analysis generally
results in integral expressions for which
analytic solutions are not possible. In such cases
numerical techniques such as Monte Carlo simulations provide an
efficient means of computing the bias. 
In order to utilize such methods, though, it is important to have a solid
understanding of the physical origin of Malmquist bias.  There are
three key points to keep in mind:
\begin{enumerate}
 \item Malmquist bias is a result of {\it observational uncertainty}, and 
all uncertainties contribute to the bias.  Only if a distance is
measured with no uncertainty will there will be no bias. 
 \item Malmquist bias is a {\it geometric effect}.  When a solid angle of 
sky is observed, the surface area covered by that solid angle increases
with distance, so a greater spatial volume is seen at greater distance.
This effect is equivalent, all other factors being equal, to giving
greater weight to more distant values and hence biasing the results.  
This makes the error envelope non-Gaussian about the observed
value, being instead skewed towards greater distances.
 \item When we try to measure the distance to an object, the Malmquist bias
associated with that observation is dependent on {\it the spatial
distribution of similar objects which might also have been observed}.
This means that the bias in a distance measurement cannot be
determined from information about the observed object alone but must 
also depend on the a priori spatial distribution of the sample from 
which the object was drawn.
As an example, consider a sample of
objects which is comprised of all Galactic G stars within a given solid 
angle regardless of brightness. Imagine that one star in the sample is chosen
at random and the
distance to this star is measured. By itself, the geometric effect would 
imply that it is most probable that this object is infinitely far away,
since the volume seen increases with distance. We know this cannot be true, 
however, because the distribution of Galactic G stars does not extend to infinity. 
Consequently, the most probable distance to the star will lie within the Galaxy.   
\end{enumerate}

	To summarize, Malmquist bias is a geometric effect. The
three-dimensional geometry of space alone tends to make the most
probable value of the true distance greater than the measured, raw
distance.  The parent distribution of objects, is also
important. It can act to increase, mitigate, or even reverse the sign
of the bias.  None of this matters though unless there is error. If
you know exactly where something is, then there can be no bias. 

\section{Malmquist Bias and the Virgo Cluster}
	
	So far the discussion has been rather general. Now, let us
return to Virgo.  
To determine the bias in the distance to Virgo, it is important to
 carefully consider each of the factors
which affect the bias. In particular, there is bias only if there is
observational error. What are the sources of error in the Key Project
distance measurement to Virgo?   Further, the bias depends on the parent
distribution of objects in space from which Virgo was selected.
 What is the parent distribution of ``Virgo-like" clusters?  For that
matter, what {\it is} a ``Virgo-like" cluster? We consider each of these items
in turn.

\subsection{Sources of Error}

	There are two distinct sources of error in the Key Project
distance to Virgo. One is the total magnitude error in the measured
distance modulus to M100.  The Key Project error budget lists this
value as 0.17 magnitudes (\cite{fer96}). The other source, more subtle
but ultimately more important, is the extended nature of the cluster itself.
The aim is to measure the distance to the center of Virgo, but M100 
is not necessarily located exactly at the center. All we have is a 
probability distribution for the location of M100 relative to the center.
This second source of error is much harder to correct for because
the structure of the Virgo cluster is neither simple nor well determined. 

	Much research has been devoted to understanding the structure
of the Virgo cluster (\cite{tul84}; \cite{san85}; \cite{bin85}; \cite{bin87}; \cite{fuk93}; \cite{yas96}).  It 
is known that the distribution of galaxies within the cluster is not
smooth, but rather exhibits several distinct density peaks. In
particular there are two prominent peaks on the sky centered near M87
(NGC 4486) and M49 (NGC 4472). Further, the distribution of galaxies
within the cluster varies with morphological type, the ellipticals
being more concentrated about the two peaks than the spirals
(\cite{san85}). Fortunately the cluster does appear to be regular
enough to allow the angular distribution of spirals to be modeled by an
exponential density profile. Shaya and Tully observed a projected
angular profile of spirals in the cluster with a scale length of $\theta=3.1
\pm0.5$ degrees centered on the M87 group, with which M100 is believed
to be associated (\cite{tul84}). A similar  analysis by Binggeli,
Sandage, and Tammann, which excluded galaxies below $\delta=9^o$ to avoid
the M49 group, yielded a value of $\theta=3.3$ degrees (\cite{bin87}).  This is the
value used in the present simulations.

	The line of sight profile of the cluster is less certain.
Several groups have observed that the Virgo 
spiral B-band Tully-Fisher relation is broader than normal 
(\cite{pie88}; \cite{fuk93}; \cite{yas96}).  If this excess dispersion is due primarily
to the depth of the cluster, as opposed to intrinsic scatter, this 
would indicate that the cluster has greater line of sight depth than width
on the sky. 

  	Insight into the depth profile of the cluster can be gained from 
the data of Fukugita et al. (\cite{fuk93}; \cite{yas96}). These authors plot the
density of spiral galaxies as a function of TF distance and postulate from this
that in fact the cluster is filamentary with the major axis nearly aligned 
with the line of sight.  
However, it is not immediately clear from their data how extended Virgo is along the line of
sight, as the points which they plot represent a 
convolution of the true depth profile of the cluster with a Gaussian
magnitude error from the Tully-Fisher relation.

	An attempt can be made to model this convolution and determine the 
effective scale of the cluster along the line of sight if a form for the
density distribution is assumed. As has been noted above, the distribution
of spirals projected on the sky is observed to be well fit by an
exponential profile of the form $e^{-|s|/a}$, where $s$ is the projected
distance from the center of the cluster and $a$ is the effective scale 
length. We will assume that the line of sight density distribution of the
cluster can also be modeled by an exponential function, but will not assume
that the line of sight scale length, $b$, is necessarily the same as the
scale length, $a$, projected on the sky. These scale lengths will be used to 
define an ellipsoidal model for the cluster; surfaces of
constant density will be described by ellipsoids with eccentricity
\begin{equation}
 e^2=1-\left(\frac{a}{b}\right)^2.
      \label{eq:eccent}
\end{equation}

In order to model the convolution,
it is necessary to know the intrinsic magnitude error associated with the
Tully-Fisher relation.
The dispersion of the Tully-Fisher relation for the
calibration galaxies of Fukugita et al. is $\sigma=0.29$ mag (\cite{yas96}). If it is
assumed that the intrinsic scatter in Virgo is equivalent to that of
the calibration galaxies, this implies a scale length 
$b=3.5\pm0.5$ Mpc and a TF distance of 14.4 Mpc to the cluster center
(see Figure 1a). If instead the dispersion is as great as 0.4 mag,
then the scale length is $3.0\pm0.5$ Mpc (see Figure 1b).  It is
unlikely that the intrinsic scatter is greater than 0.4 mag, so 
the probable value of the scale length lies in the range 2.5-4.0 Mpc.
As can be seen, even after deconvolving the data there remains
considerable uncertainty in the depth profile of the cluster. 
This is one of the sources of uncertainty in calculating the Malmquist correction
to the cluster distance.

	To summarize, the assumed model for the space density of spiral
galaxies within Virgo in cluster-centric coordinates is
\begin{equation}
\rho(r,\phi)=e^{-r/l(\phi)},	\label{eq:spirals}
\end{equation}
where $l(\phi)$ is defined by the expression
\begin{equation}
l(\phi)=\frac{a}{\sqrt{1-e^2\cos(\phi)^2}}.
\end{equation}
Here $\phi$ is the azimuthal angle of the cluster coordinate system,
and $e$ is the eccentricity of the ellipsoidal model of the cluster. The angle $\phi$ is
defined such that $\phi=0$ along the line of sight on the near side of the cluster
and $\phi=\pi$ along 
the line of sight on the far side. The eccentricity
is defined in equation (\ref{eq:eccent}),
where $b$ is the line of 
sight scale length and $a$ is the scale length projected on the sky. In terms
of observables, $a=r_c\tan(\theta)$, where $r_c$ is the distance from us to the center of
the cluster, and $\theta$ is the angular scale length projected on the sky. Current
information on the structure indicates $b=2.5-4.0$ Mpc, and $\theta=3.3^{o}$.
For $r_c\approx 17$ Mpc, this would imply $a\approx 1.0$ Mpc and $\frac{b}{a}\approx 
2.5-4$, a ratio similar to that of an E6 or E7 elliptical galaxy.

\begin{figure}
\plotone{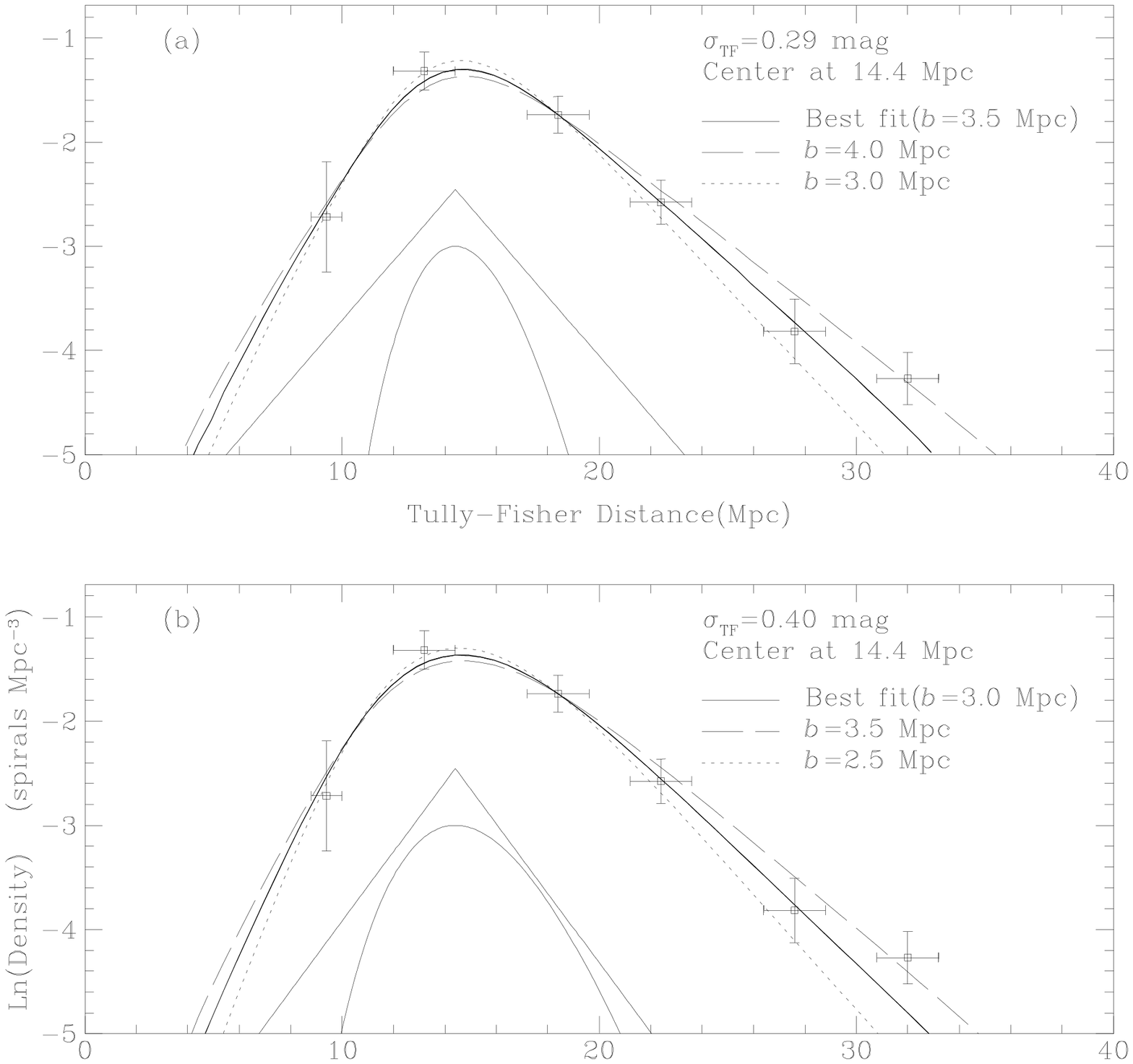}
\figcaption[profile.eps]{Two models for the spatial density of
 spiral galaxies towards Virgo as a function
of raw TF distance (Fukugita et al 1993). The heavy solid curves are the
 best fit cases.
The dashed and dotted curves are for values of {\it b} which are 0.5 Mpc
greater or less than
the best fit, respectively. The two solid curves beneath the fit show the form of the
intrinsic distribution and Gaussian magnitude error which were convolved.
 \label{fig1}}
\end{figure}

\placefigure{fig1}
	
\subsection{The Parent Distribution of Virgo-like Clusters}

	The second key issue is the parent
distribution of clusters from which Virgo was drawn. It appears that Virgo was
chosen by the Key Project because it is the nearest large cluster to the 
Milky Way.  Clearly, only one cluster has that status, and it has exactly one
distance, not a distribution of distances.  Nevertheless, there is a probability
distribution of potential distances from which the real distance was drawn. This
is the relevant $r^2 n(r)$ in equation (\ref{eq:jntprob}). 

	To model $n(r)$, we consider an ensemble of statistically identical 
universes\footnote{Alternatively, we could consider all pairs in our universe
of Milky Way-like galaxies and their nearest large clusters.}. For simplicity, we
assume that the populations of Milky Way-like galaxies and large clusters are
independently distributed at random with no mutual correlation. We also assume that
the distribution of clusters is isotropic. If we place ourselves 
in any one Milky Way, the distribution $n(r)$ then depends only on the mean number 
density of large clusters, assuming that the average density is independent of $r$.
If that density is high, the nearest cluster will be
close by, and $n(r)$ will peak at a low value of $r$.  If that density is low, 
the nearest cluster will be distant, and $n(r)$ will peak far away.  Analytically, 
this function can be derived using Poisson statistics. This yields 
\begin{equation}
r^2 n(r)=4\pi r^2 n_0 e^{-4\pi r^3 n_0/3},
\end{equation}
where $n_0$ is the average cluster density.

	The next question is: what is the density of clusters in 
the universe? More specifically, what is the density of clusters
which are at least as rich as Virgo\footnote{We assume that the Key Project would have 
been content to select any cluster as rich or richer than Virgo.}?  This question
turns out to have a non-trivial effect on the magnitude of the
bias.  
Scaramella et al. carried out a detailed study of the Abell and ACO cluster catalogs 
(\cite{sca91}). 
After correcting for selection effects, they determined that the density of clusters
of richness class 0 or greater was $n_0=1.46\pm0.12\times10^{-5} h^{3}$ Mpc$^{-3}$ for the Abell
catalog and $n_0=2.35\pm0.15\times10^{-5} h^{3}$ Mpc$^{-3}$ for the ACO catalog. The 
difference in these two numbers is attributed to several factors, including greater 
completeness of the ACO catalog, more spurious detections in the ACO catalog due to a 
different method of cluster detection, and potentially real effects since the two 
catalogs sample different regions of the sky.  Although these density values differ
by roughly 60\%,  they do provide a rough range within which the true density
of Abell type clusters likely lies.

Virgo is
not an Abell or ACO cluster, but it is generally believed to be
comparable to Abell clusters of richness class 0
(\cite{hec81}; \cite{vig89}). Consequently, the Abell and ACO densities
can serve as 
reasonable estimates of the density of clusters of Virgo richness or greater in
the local universe. 

\subsection{Further Constraints from Velocity Information}

        The previous two sections describe the basic geometrical situation which must be
modelled.  However, we also have velocity information which should be
utilized to make modelling of the bias more realistic.

	In the basic model, it is assumed that the nearest cluster can lie at any
distance, subject to a given probability distribution. The only constraint on the
resulting bias is that the observed distance conform to the distance measured to M100.
In reality, the simulated clusters must also conform to velocity data available for
the cluster and M100. 

	First consider the velocity of the Virgo cluster. If we have some {\it a priori} 
knowledge of the possible range of values for $H_0$, then the velocity of the cluster
should be used to restrict the range of distances at which the cluster may lie. There 
appears to be some consensus in the literature that the value of $H_0$ lies in the 
range $90>H_0>50$ km s$^{-1}$ Mpc$^{-1}$ (\cite{ken95}).  Also, the recessional velocity of 
Virgo relative to the Local Group is 1396$\pm$96 km s$^{-1}$ (\cite{huc95}). Taking $1300
<v_{Virgo}<1500$ km $s^{-1}$ and $90>H_0>50$ km s$^{-1}$ Mpc$^{-1}$, we can roughly restrict
$r$, the true distance to the center of the Virgo cluster, to be $14.4<r<30$ Mpc. This 
restriction has very little effect at the upper end of possible cluster distances, but the constraint 
that $r> 14.4$ Mpc excludes a significant fraction of candidate clusters (see Figure \ref{fig2}). 
Consequently, the bias is greater here than in the basic model. 

	Further, an attempt can be made to utilize information on the velocity of M100
relative to the cluster to constrain its distance relative to Virgo's. Yasuda et al. 
plot heliocentric velocity versus Tully-Fisher distance for spirals in Virgo (\cite{yas96}).
In this plot, very little correlation is seen between these parameters due to virialization of the 
core. However, no spiral galaxy with a heliocentric velocity greater than 1000 km s$^{-1}$ 
is observed at a Tully-Fisher distance less than 10 Mpc. M100 has a heliocentric velocity 
of 1580 km s$^{-1}$.  It was noted previously that the best deconvolution of this data 
set gives a Tully-Fisher distance of 14.4 Mpc to the center of the cluster. We can 
conservatively use this information to place a bound on the ratio of the distance to 
M100 relative to the distance to the center of Virgo. Specifically, this information 
implies that $d_{M100}> \frac{2}{3} d_{Virgo}$.

\section{Monte Carlo Simulations}
\subsection{Method}

	Given the above information on Virgo, a Monte Carlo routine
was constructed to numerically evaluate the Malmquist bias in the
distance to the cluster center. The center of the cluster was defined
to be the peak of the cluster's spiral density distribution. 

  	The logic behind the routine was as follows: First a true
distance to the center of a cluster was picked at random based on the
assumptions regarding the distribution of Virgo-like clusters. Next a
galaxy within the cluster was selected at random based on the assumed
distribution of spirals in the cluster, as given in equation (\ref{eq:spirals}).
The distance to
this galaxy, $r_g$, was then calculated and a Gaussianly distributed
magnitude error was added corresponding to the measurement error in the
actual observation. This gave a measured, raw distance, $d_o$. If $d_o$
was equal to the Key Project distance of 16.1 Mpc to within 0.5
Mpc and the galaxy was at a projected separation of less than 6 degrees
from the center of the cluster, then the datum was considered consistent. If 
consistent, the measured distance, $d_o$, and actual distance to the 
cluster center, $r_c$, were kept.  Otherwise the trial was thrown out.  
This procedure was repeated until a set number of successful iterations, 
typically 20,000, were completed. 

	The Malmquist bias to the cluster was then computed.
If $R$ is defined to be the unbiased true distance to the cluster,
then $R=(\Sigma r_{c,i})/N$, where $N$ is the number of trials. 
Likewise, the mean measured distance to the cluster is 
$D_o=(\Sigma d_{o,i})/N$. The true distance is related to the observed
distance by the equation 
\begin{equation}
R=(1+B)D_o,
\end{equation}
where $B$ is the Malmquist bias. Rewriting this equation in the form
\begin{equation}
B=\frac{R}{D_o} - 1,
\end{equation}
 the bias can be readily calculated.

\subsection{Simulations}

	Simulation of the bias was approached in three stages in order to illustrate
the effect of each additional level of information. First, 
simulations were run for the basic model, ignoring velocity data. Next velocity 
information for the cluster was included, effectively modifying $n(r)$, and in the
third stage the velocity information for M100 was used to further constrain $P(r|D)$.

	As mentioned, uncertainty exists regarding the line of sight
distribution of spirals in the cluster.  With reasonable confidence
it can be assumed that the effective line of sight scale length, {\it b}, of the
cluster is between 2.5 and 4.0 Mpc. Monte Carlo simulations were run
for a range of scale lengths, and for both Abell and ACO densities. 
The velocity of Virgo was taken to be 1,396 km s$^{-1}$.

\begin{figure}
\plotone{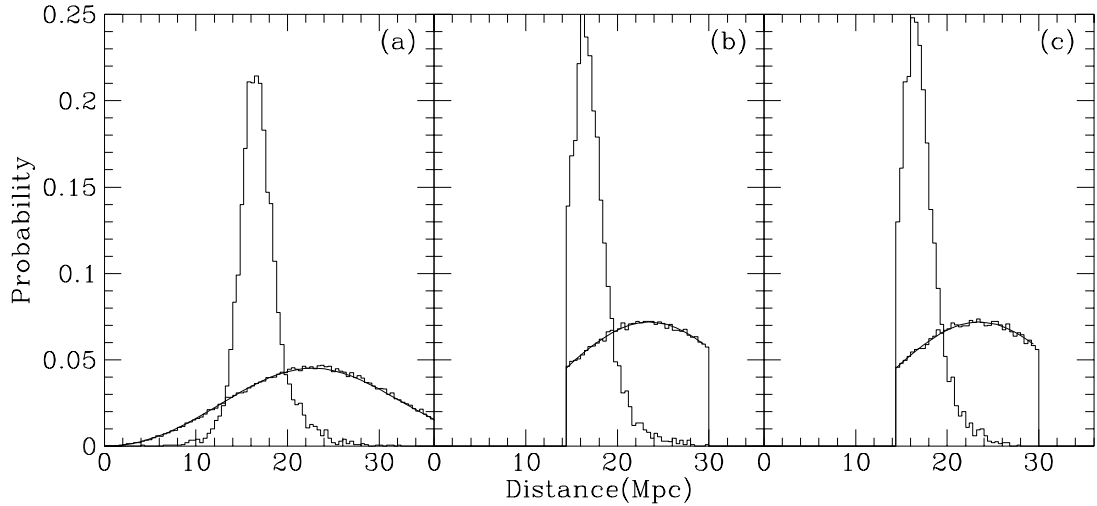}
\figcaption[prob.eps]{The lower histogram in each frame shows the 
distribution of distances to the nearest cluster in the Monte Carlo
simulations. The solid curve beneath it is the analytic function for
this distribution. The upper histogram is the distribution of distances
to the nearest cluster which produced the observed distance to M100 in
the simulations for ACO density, $b$=3.0.  Figures 2(a),2(b), and 2(c)
are for the basic model, basic model plus $H_0$ cutoff, and full model, 
respectively. All curves are normalized to unit probability.\label{fig2}}
\end{figure}

 	A plot of the underlying cluster probability distribution and the 
distribution of clusters which satisfied the selection criteria is shown for all 
three models in Figure \ref{fig2}.
\placefigure{fig2}

\subsubsection{Basic Model}
\begin{figure}[h] 
\vspace{3in}
        \begin{minipage}[t]{3in}
        \epsfxsize=3in
        \epsfysize=3in
        \epsfbox{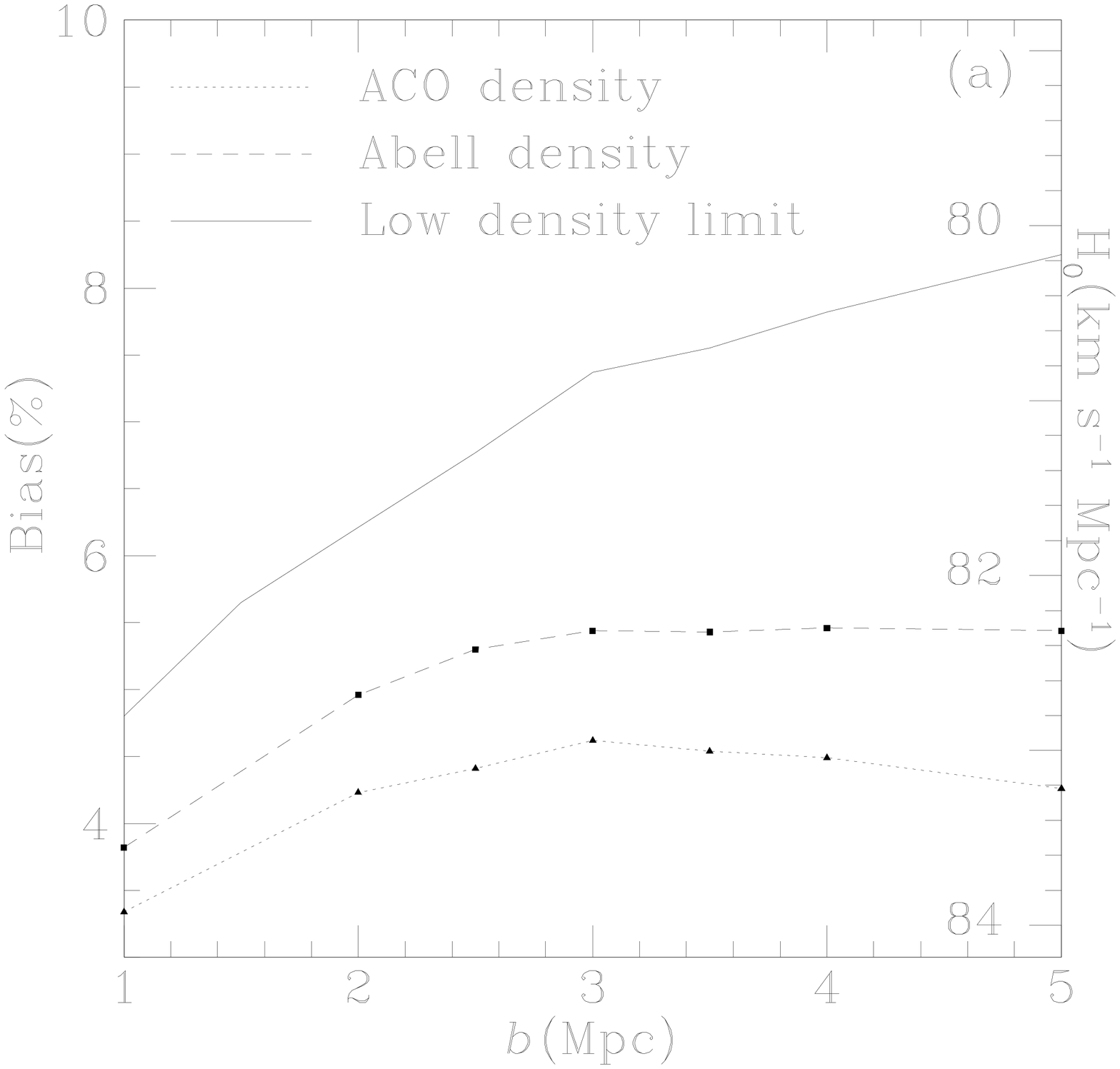}
	\end{minipage}\hfill
	\begin{minipage}[t]{3in}
        \epsfxsize=3in
        \epsfysize=3in
        \epsfbox{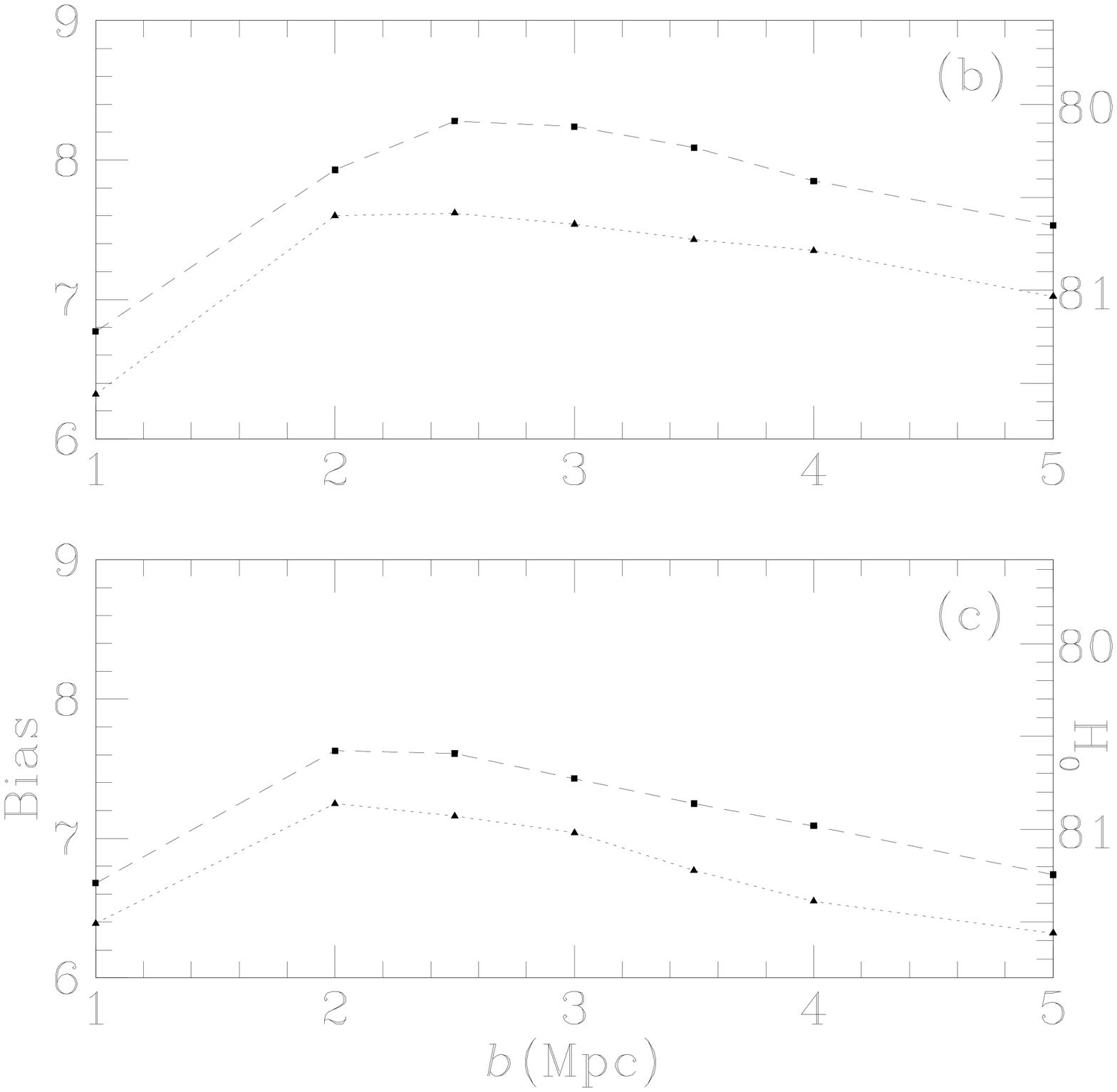}
	\end{minipage}
\figcaption[bias.a.eps]{Bias as a function of the depth profile of Virgo
for Abell and ACO density. For the basic model, the low density limiting case
is also shown. Figures 3(a),3(b), and 3(c) are for the basic model, basic
model plus $H_0$ cutoff, and full model, respectively.\label{fig3}}
\end{figure}

	Results for the basic model can be seen in Table \ref{tbl-1} and Figure \ref{fig3}. 
That the bias increases with increasing {\it b} is intuitively plausible.
Larger {\it b} increases the cluster depth along the line of sight,
and thus the relative uncertainty in the distance to the cluster as
determined by M100. This may be thought of as an additional error, and
hence the bias increases. 
That the bias approaches a constant value for large
scale lengths is less intuitive. This is due to the behavior of $n(r)$. Remember that
Virgo is the nearest cluster. For large $r$, the probability that a cluster at distance
$r$ is the nearest decreases rapidly. Consequently, although increasing $b$ allows clusters
farther away to appear closer, at large $r$ the parent population $n(r)$ permits only a
negligible contribution to the bias. Least intuitive is that the magnitude of the bias
peaks before declining to a constant value for larger scale lengths.  This effect is 
associated with $n(r)$. The magnitude of the bias decreases once the
scale length is great enough to allow significant sampling of the region 
beyond the peak of  $r^2 n(r)$.

For comparison, simulations were also run in which any cluster within 120 Mpc could
be Virgo, regardless of whether it was the nearest cluster, as long as the simulation
produced an observed distance of $16.1\pm0.5$ Mpc.  This corresponds to the
low density limit of the nearest cluster case. 
In the low density case, the magnitude of the bias is greater and
varies more with scale length. Note also that here the bias does not decline or approach a 
constant value for large $b$, but rather continues to increase (Figure \ref{fig3}(a)).

 \subsubsection{Basic Model + Virgo Velocity Information}
	These simulations include the constraint $50<H_0<90$ km s$^{-1}$ Mpc$^{-1}$. The results can be
seen in Table \ref{tbl-1} and Figure \ref{fig3}(b). It should be noted that the magnitude of the bias is 
quite sensitive to exactly where the high $H_0$ cutoff is made. For example,
for $b=3.0$ and ACO density, changing the minimum allowed distance from 14.4 
to 13.7 (i.e. changing the maximum $H_0$ from 90 to 95) lowers the bias from 7.5\% to 6.4\%.
For plausible scenarios with $d_{cutoff}=14.4$, this model yields a bias of 7.4-8.3\%.

\subsubsection{Full Model}
	This adds the second constraint that $d_{Virgo}<\frac{3}{2} d_{M100}$.
The results can be seen in Table \ref{tbl-1} and Figure \ref{fig3}(c).
The form of the bias as a function of scale length is very similar to the previous model, only 
with a smaller bias. The primary difference is that the bias decreases more rapidly now
with increasing $b$ since a greater fraction of distant clusters are now excluded.
Plausible parameters now yield a bias of 6.5-7.6\%.
\placetable{tbl-1}
\placefigure{fig3}

\section{Discussion}

	 We have shown that, with present data, Malmquist bias noticeably alters the
most probable distance to the Virgo cluster. Unfortunately, it is not possible 
to determine the precise magnitude of the bias. In the basic model,
this is due to two key uncertainties: 
i) uncertainty in the spiral galaxy depth profile in the cluster, and
ii) uncertainty as to the density and distribution of clusters. Of the
two, the second leads to a greater uncertainty in the bias.  In the models which
include velocity information, it is seen that additional information regarding the
cluster changes $n(r)$ and $P(r|D)$ and hence the magnitude of the bias. This additional
knowledge leads to a more realistic estimation of the bias; however, it also contributes
uncertainty in the magnitude of the bias since this knowledge is incomplete.

	Given present knowledge, the magnitude of the bias is likely 6.5-8.5\%. For a raw
distance of 16.1 Mpc, this implies an unbiased true distance of 17.2 to 17.4 Mpc.  $H_0$
likewise decreases from a raw value of 87 km s$^{-1}$ Mpc$^{-1}$ to an
unbiased value of 80-82 km s$^{-1}$ Mpc$^{-1}$.  In particular, if
the density of clusters is taken to be comparable to the ACO
density in the full model, with $b=3.5$ Mpc, this yields d$=17.2\pm1.9$ Mpc 
and $H_0=81\pm 11$ km s$^{-1}$ Mpc$^{-1}$. 
This is one of the scenarios most consistent with the data, assuming
that the intrinsic dispersion of the TF relation for Virgo is
comparable to that of field galaxies. 
	Finally, it should be realized that the bias computed here includes only the bias 
associated directly with the Cepheid-based determination of the distance 
to the Virgo cluster. In fact, the calibration of the Cepheid P-L relation is dependent on
the distance to the Large Magellanic Cloud. Bias in the distance to the LMC would lead
to additional bias in the distance to Virgo. The effect of this bias has not been 
included in this paper.

\section{Conclusions}
	The Malmquist bias associated with the H$_0$ Key Project 
determination of the distance to the Virgo cluster is of order 6-8\%, which is
significant. However, this particular bias 
correction is only of temporary interest, as Virgo will soon cease to be the lynchpin of
the Key Project.  Of more lasting interest, this exercise serves to illustrate 
several general issues which often arise when considering the effect of Malmquist bias on
astronomical distance measurements.

	One issue is the subtle effect on the bias of exactly how an object is defined. The
definition of the object defines the parent population from which the object
is drawn, $n(r)$, and hence changes the bias. In the case of Virgo, there is a clear
change in the bias resulting from changing the parent population from {\it all}
large clusters to the {\it nearest} large cluster\footnote{Note that even this
treatment of $n(r)$ falls short because it does not include a probable positive 
statistical correlation between the Milky Way and Virgo-like clusters due to the 
galaxy-cluster correlation function. However, our aim is to be illustrative rather than
exact, so we have omitted this.}.
	Similarly, the exact bias is sensitive to the spatial model of the cluster, 
$\rho(r,\phi)$. The extended nature of the cluster introduces uncertainty which 
contributes to the bias in the same fashion as other observational
errors. The more extended the cluster is along the line of sight, the greater the bias. 

	Another important fact illustrated here is that Malmquist bias
is present when the distance to a single object is measured. Most 
published astronomical distances are to single objects. They are essentially {\it
all} biased, most in such a way that the measured distance is {\it too
small}.  This is as true for parallax measurements to stars as for extragalactic measurements
\footnote{In the field of astrometry this bias is referred to as the Lutz-Kelker effect, and
is generally of much greater magnitude than in our present case.} (\cite{lut73}; \cite{han79}).

	The implications of this are particularly important with regard to ``distance
ladders". With each additional step in such a ladder, the total uncertainty increases.
For instance,
for Gaussian magnitude errors, the uncertainty increases in quadrature as
$\sigma_{tot}^{2}=\Sigma\sigma_{i}^{2}$. Thus, for the simple case of constant 
$n(r)$ and $\sigma_{i}$, the bias increases linearly with the number of 
steps (Equation \ref{eq:lyndbell}). 
In light of this fact, it is clearly important to minimize the number of steps,
and to treat properly any bias in those steps that are retained.

\section{Acknowledgments}
AG would like to thank everyone in the office for letting him bounce ideas off of them. He
also acknowledges the NSF for support of this work under a Graduate Research Fellowship.
SF would like to thank Jeff Willick and Avishai Dekel for their discussions regarding
Malmquist bias.  We both wish to thank the anonymous referee for the excellent suggestion that
we include velocity constraints in the models.

\placetable{tbl-2}

\clearpage

\begin{deluxetable}{cccccccc}
\tablecaption{Results of Monte Carlo simulations. \label{tbl-1}} 
\tablewidth{0pt}
\tablehead{
\colhead{$b$}  &
\colhead{$B$}  & \colhead{$d_{Virgo}$} &  \colhead{$H_0$} &
\colhead{} & \colhead{$B$} & \colhead{$d_{Virgo}$} & \colhead{$H_0$} 
}

\startdata
\multicolumn{4}{c}{Abell Density} & \multicolumn{4}{c}{ACO Density}   \nl
\hline
\hline
\multicolumn{8}{c}{Basic Model} \nl
2.5  & 5.5 & $17.0\pm2.5$ & $82\pm13$ &  & 4.4 & $16.8\pm2.4$ & $83\pm13$ \nl
3.0  & 5.4 & $17.0\pm2.6$ & $82\pm14$ &  & 4.6 & $16.8\pm2.4$ & $83\pm13$ \nl
3.5  & 5.4 & $17.0\pm2.6$ & $82\pm14$ &  & 4.5 & $16.8\pm2.5$ & $83\pm14$ \nl
4.0  & 5.5 & $17.0\pm2.7$ & $82\pm14$ &  & 4.5 & $16.8\pm2.4$ & $83\pm13$ \nl
\hline 
\multicolumn{8}{c}{Basic Model plus $H_0$ Cutoff}\nl
2.5  & 8.3 & $17.4\pm2.2$ & $80\pm12$ &  & 7.6 & $17.3\pm2.1$ & $81\pm11$ \nl
3.0  & 8.2 & $17.4\pm2.3$ & $80\pm12$ &  & 7.5 & $17.3\pm2.1$ & $81\pm11$ \nl
3.5  & 8.1 & $17.4\pm2.3$ & $80\pm12$ &  & 7.4 & $17.3\pm2.1$ & $81\pm11$ \nl
4.0  & 7.9 & $17.4\pm2.2$ & $80\pm12$ &  & 7.4 & $17.3\pm2.1$ & $81\pm11$ \nl
\hline
\multicolumn{8}{c}{Full Model}\nl
2.5  & 7.6 & $17.3\pm2.0$ & $81\pm11$ &  & 7.2 & $17.3\pm1.9$ & $81\pm10$ \nl
3.0  & 7.4 & $17.3\pm2.0$ & $81\pm11$ &  & 7.0 & $17.2\pm1.9$ & $81\pm11$ \nl
3.5  & 7.3 & $17.3\pm2.0$ & $81\pm11$ &  & 6.8 & $17.2\pm1.9$ & $81\pm11$ \nl
4.0  & 7.0 & $17.2\pm1.9$ & $81\pm11$ &  & 6.6 & $17.2\pm1.8$ & $81\pm10$ \nl

\enddata
\tablecomments{$b$: Line of
sight scale length(Mpc), $B$: Malmquist bias(\%),
$d_{Virgo}$: Bias-corrected Distance to Virgo(Mpc), $H_0$: Bias-corrected
$H_0$(km s$^{-1}$ Mpc$^{-1}$) }
\end{deluxetable}

\clearpage

\begin{deluxetable}{lrrr}
\tablecaption{Key Project and Malmquist corrected data. \label{tbl-2}} 
\tablewidth{0pt}
\tablehead{
\colhead{} & \colhead{} & \colhead{Abell Density} & \colhead{ACO Density} \nl
\colhead{} & \colhead{Key Project Value}  & \colhead{Corrected Value}&\colhead{Corrected Value}
}

\startdata
Virgo Recessional Velocity&  $1396\pm96$ km s$^{-1}$& \nodata & \nodata \nl
Distance to M100&  $16.1\pm1.3$ Mpc& \nodata & \nodata \nl
Distance to Virgo& $16.1\pm3.5$ Mpc& $17.3\pm 2.0$ & $17.2\pm 1.9$ \nl
$H_0$&  $87\pm20$ km s$^{-1}$ Mpc$^{-1}$& $81\pm11$ & $81\pm11$  \nl
\tablecomments{The error in the Key Project distance to Virgo corresponds to a 20\% Gaussian
uncertainty added in quadrature to the uncertainty in the distance to M100. Also, the value of $H_0$
in the Key Project column is not directly from Ferrarese et al. Rather, it was calculated here
from the information given in that paper.}
\enddata
\end{deluxetable}

\clearpage

\clearpage

\clearpage

\clearpage



\begin{thebibliography}{}

\bibitem[Bachall 1988]{bac88} Bachall, N.A.,
    Annual Reviews of Astronomy and Astrophysics, 26, 631
\bibitem[Binggeli, Sandage, \& Tammann 1985]{bin85} Binggeli, B., Sandage, A.,
    \& Tammann, G. 1985, \aj, 90, 1681
\bibitem[Binggeli, Sandage, \& Tammann 1987]{bin87} Binggeli, B., Sandage, A.,
    \& Tammann, G. 1987, \aj, 94, 251
\bibitem[Ferrarese et al. 1996]{fer96} Ferrarese, L., et al. 1996, \apj,
    464, 568
\bibitem[Foqu\`e et al. 1990]{foq90} Foqu\`e, P., Bottinelli, L.,
    Gougenheim, L., \& Paturel, G. 1990,\apj, 349, 1

\bibitem[Freedman et al. 1994]{fre94} Freedman, W., et al. 1994, Nature,
    371, 757
\bibitem[Fukugita, Okamura, \& Yasuda 1993]{fuk93} Fukugita, M.,
    Okamura, S., \& Yasuda, N. 1993, \apj, 412, L13
\bibitem[Hanson 1979]{han79} Hanson, R. 1979, \mnras, 186, 875
\bibitem[Heckman 1981]{hec81} Heckman, T.M. 1981, \apj, 250, L59
\bibitem[Huchra 1995]{huc95} Huchra, J.P. 1995, The MSSSO Heron Island Workshop
on Peculiar Velocities in the Universe, http://msowww.anu.edu.au/~heron/Huchra/huchra.html
\bibitem[Kennicutt, Freedman, \& Mould 1995]{ken95} Kennicutt, R.,
    Freedman, W., \& Mould, J. 1995, \aj, 100, 1475
\bibitem[Landy and Szalay 1992]{lan92} Landy, S., \& Szalay A. 1992,
    \apj, 391, 494 
\bibitem[Lutz and Kelker 1973]{lut73} Lutz, T., \& Kelker, D. 1973, \pasp, 85, 573
\bibitem[Lynden-Bell et al. 1988]{lyn88} Lynden-Bell, D., et al. 1988, \apj, 326, 19
\bibitem[Mould et al. 1995]{mou95} Mould, J., et al. 1995, \apj, 449, 413
\bibitem[Malmquist 1924]{mal24} Malmquist, K.G. 1924, Medd. Lund Astron. Obs.,
    Ser. II, No. 32, 64
\bibitem[Pierce and Tully 1988]{pie88} Pierce, M., \& Tully, R. 1988, \apj,
    330, 579
\bibitem[Sandage, Binggeli, \& Tammann 1985]{san85} Sandage, A., Binggeli, B.,
    \& Tammann, G. 1985, in The Virgo Cluster, ed. O. Richter \& B. Binggeli 
    (Garching:ESO), 181
\bibitem[Scaramella et al. 1991]{sca91} Scaramella, R., Zamorani, G., Vettolani,
    G., Chincarini, G., \aj, 101, 342
\bibitem[Strauss and Willick 1995]{str95} Strauss, M., \& Willick, J. 1995,
    Physics Reports, 261, 271
\bibitem[Tully and Shaya 1984]{tul84} Tully, R., \& Shaya, E. 1984,  \apj,
    281, 31
\bibitem[Vigroux, Boulade, \& Rose 1989]{vig89} Vigroux, L., Boulade, O., Rose, 
    J. A. 1989, \aj, 98, 2044
\bibitem[Willick 1991]{wil91} Willick, J. 1991, PhD. thesis,
    University of California, Berkeley
\bibitem[Yasuda, Fukugita, \& Okamura 1996]{yas96} Yasuda, N., Fukugita, M.,
    \& Okamura, S. 1996, submitted to \apj
\end{thebibliography}
\end{document}